\documentclass[twocolumn,showpacs,preprintnumbers,amsmath,amssymb,superscriptaddress]{revtex4}

\usepackage{graphicx,epsf}
\usepackage{dcolumn}
\usepackage{bm}
\usepackage{hyperref}
\usepackage{latexsym}

\newcommand{\xsize}{\epsfxsize=7.0cm}

\begin{document}

\title{Fractal Dimension of 3-Blocks in 4d, 5d, and 6d\\ Percolation Systems}

\author{Gerald \surname{Paul}}
\email{gerryp@bu.edu}
\author{H. Eugene \surname{Stanley}}

\affiliation{Center for Polymer Studies and Department of Physics,
             Boston University, Boston, MA 02215}
 
\begin{abstract}

Using Monte Carlo simulations we study the distributions of the 3-block
mass $N_3$ in 4d, 5d, and 6d percolation systems. Because the
probability of creating large 3-blocks in these dimensions is very
small, we use a ``go with the winners'' method of statistical
enhancement to simulate configurations having probability as small as
$10^{-30}$. In earlier work, the fractal dimensions of 3-blocks, $d_3$,
in 2d and 3d were found to be $1.20\pm 0.1$ and $1.15\pm 0.1$,
respectively, consistent with the possibility that the fractal dimension
might be the same in all dimensions. We find that the fractal dimension
of 3-blocks decreases rapidly in higher dimensions, and estimate
$d_3=0.7\pm 0.2$ (4d) and $0.5\pm 0.2$ (5d).  At the upper critical
dimension of percolation, $d_c=6$, our simulations are consistent with
$d_3=0$ with logarithmic corrections to power-law scaling.
\end{abstract}

\maketitle

\section{Introduction}

Percolation is a classic model for disorder \cite{Ben-Avraham, Stauffer,
Bunde and Havlin}. Recently it has been recognized for bond percolation
that clusters and blobs are the $k=1$ and $k=2$ cases of $k$-connected
graphs ($k$-blocks), graphs in which all vertices are connected to every
other vertex in the $k$-block by at least $k$ independent paths
\cite{Paul,Jacobsen2002}. The fractal dimension $d_3$ of 3-blocks in 2-
and 3-dimensional percolation systems at the percolation threshold were
found to be $1.20\pm 0.1$ and $1.15\pm 0.1$, respectively \cite{Paul}.

The fact that the fractal dimensions are identical within error bars is
consistent with the possibility that $d_3$ might be independent of
dimension. The focus of this paper is to determine $d_3$ for $d=4$, 5,
and 6 using Monte Carlo simulations.

In the next section we study $d_3$ for percolation on the Cayley tree in
order to gain insight into the behavior of 3-blocks in very high
dimension. In Section \ref{SimulationMethod} we discuss the methods we
use to generate large 3-blocks in 4d, 5d, and 6d. In Section
\ref{Results} we discuss our results.

\section{Cayley Tree Results}

Percolation on the Cayley tree has been used as a model
for percolation for $d\geq 6$, the upper critical dimension of
percolation. The cluster fractal dimension and blob fractal dimension,
as well as a number of other critical exponents on the Cayley tree, are
identical to those of percolation for $d\geq 6$
\cite{Ben-Avraham,Stauffer,Bunde and Havlin}. Below we argue that for
percolation on the Cayley tree $d_k=0$ for $k\geq 3$ suggesting that
while $d_3$ may change little between $d=2$ and $d=3$, eventually $d_3$
decreases more rapidly approaching zero for $d=6$.

\begin{figure}[tbh]
\centerline{
\epsfxsize=9.0cm
\epsfclipon
\epsfbox{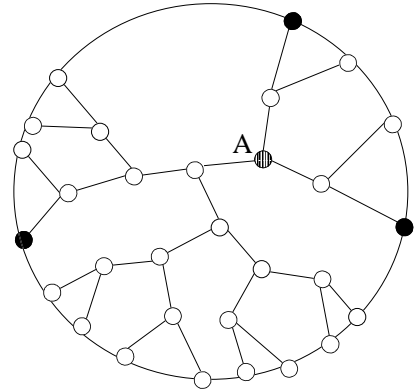}
}

\caption{Cayley tree in which each site except those on boundary
has three neighbors. We note that even when all bonds are fully occupied,
there can be only one site which is connected to three sites on the
boundary. In this example the three sites on the boundary are the filled
circles and the one site connected to them by independent paths is the
striped circle denoted A.}
\label{Cayley}
\end{figure}

To show that $d_k=0$ for $k\geq 3$ for percolation on the Cayley tree,
we make use of the concept of k-bone. Reference~\cite{Paul} generalizes
the concept of backbone by defining a $k$-bone as the set of all sites
connected to $k$ disjoint sets of points by $k$ independent paths. Thus
clusters and backbones are $k$-bones with $k=1$ and 2 respectively. For
a given $k$, the fractal dimension of $k$-bones and $k$-blocks are equal
\cite{Paul}.

To see that for percolation on the Cayley tree the fractal dimension of
a 3-bone is zero, we choose any 3 points on the boundary
(Fig.~\ref{Cayley}) and observe that there is only one site which is
connected to these points by independent paths.  This result is
independent of the size of the tree, even if the tree is fully
populated. Hence the fractal dimension is zero. Clearly this argument
holds for larger $k$ and is meaningful as long as the branching factor
in the Cayley tree is greater than or equal to $k$, and holds
independent of size.

\section{Simulation Method}
\label{SimulationMethod}
\subsection{Statistical Enhancement Method}

Randomly generated realizations in which large 3-blocks are present
become more and more rare as the system dimension increases. In fact, if
traditional techniques are used to generate realizations, for $d$ as low
as 4 the range of the values of the masses of 3-blocks created are so
small that one cannot determine $d_3$ either by finding the best
collapse of plots of the distribution of masses or by inferring $d_3$
from the slope of the power-law regime of the distributions.

To overcome this problem, we use a ``go with the winners'' method of
statistical enhancement described in Ref.~{\cite{Grassberger}.  The
basic idea of this approach in the context of a percolation cluster
growth algorithm is as follows:

\begin{itemize}

\item[{(i)}] Before we start growing a cluster, assign a value of one to
the weight $W$ of the cluster.

\item[{(ii)}] We use the Leath method to grow clusters
\cite{Leath}. While the cluster is growing, we calculate certain
properties of the state of the cluster after every interval of $n$
chemical shells of growth.

\item[{(iii)}] If certain criteria on the properties of the state of the
cluster that are described below are met, we ``clone'' the state so we
have $m$ copies (including the original) of the state, adjust $W$
accordingly to $W/m$ and continue growing each of these $m$ clones. If
these criteria are not met, simply continue growing the non-cloned
cluster.

\end{itemize}

\begin{figure}[tbh]
\centerline{
\xsize
\epsfclipon
\epsfbox{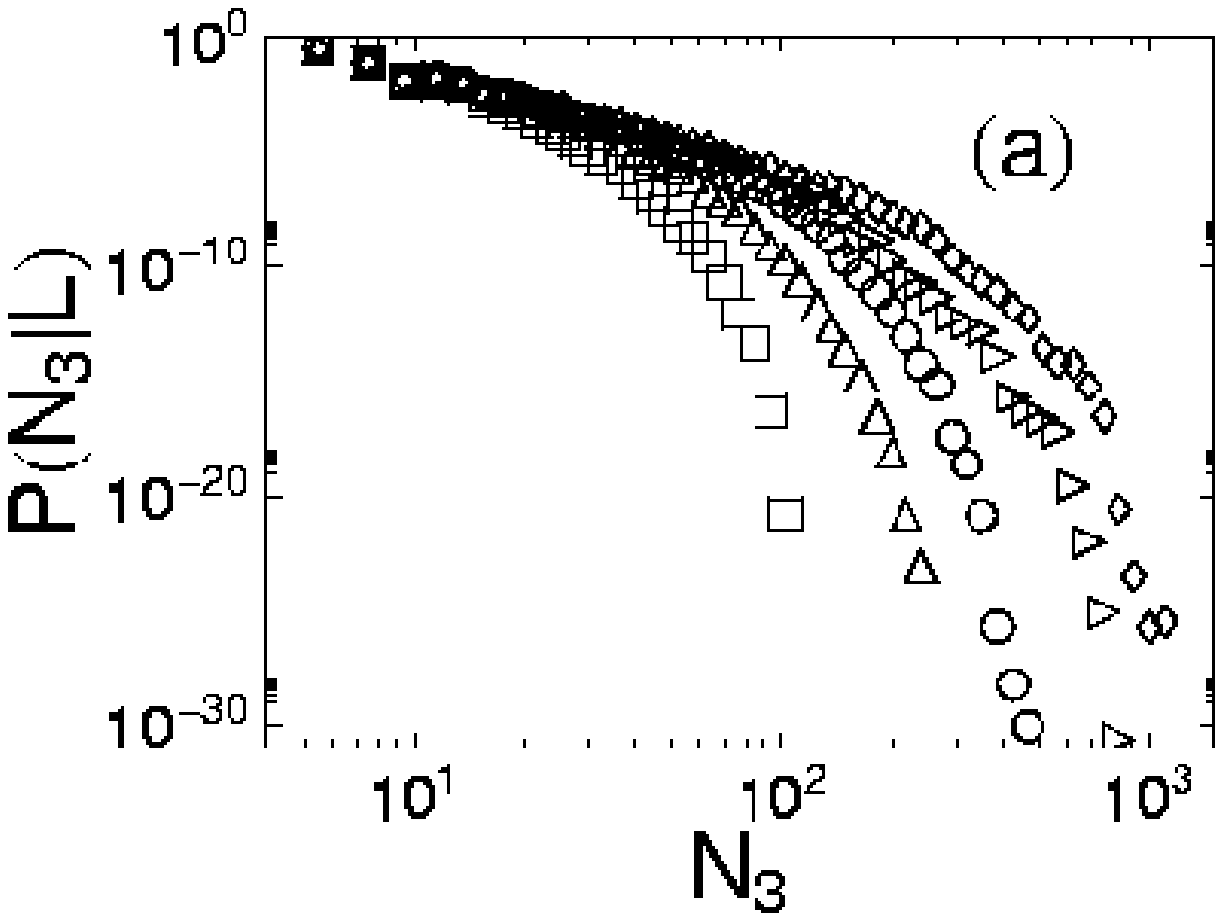}
}
\centerline{
\xsize
\epsfclipon
\epsfbox{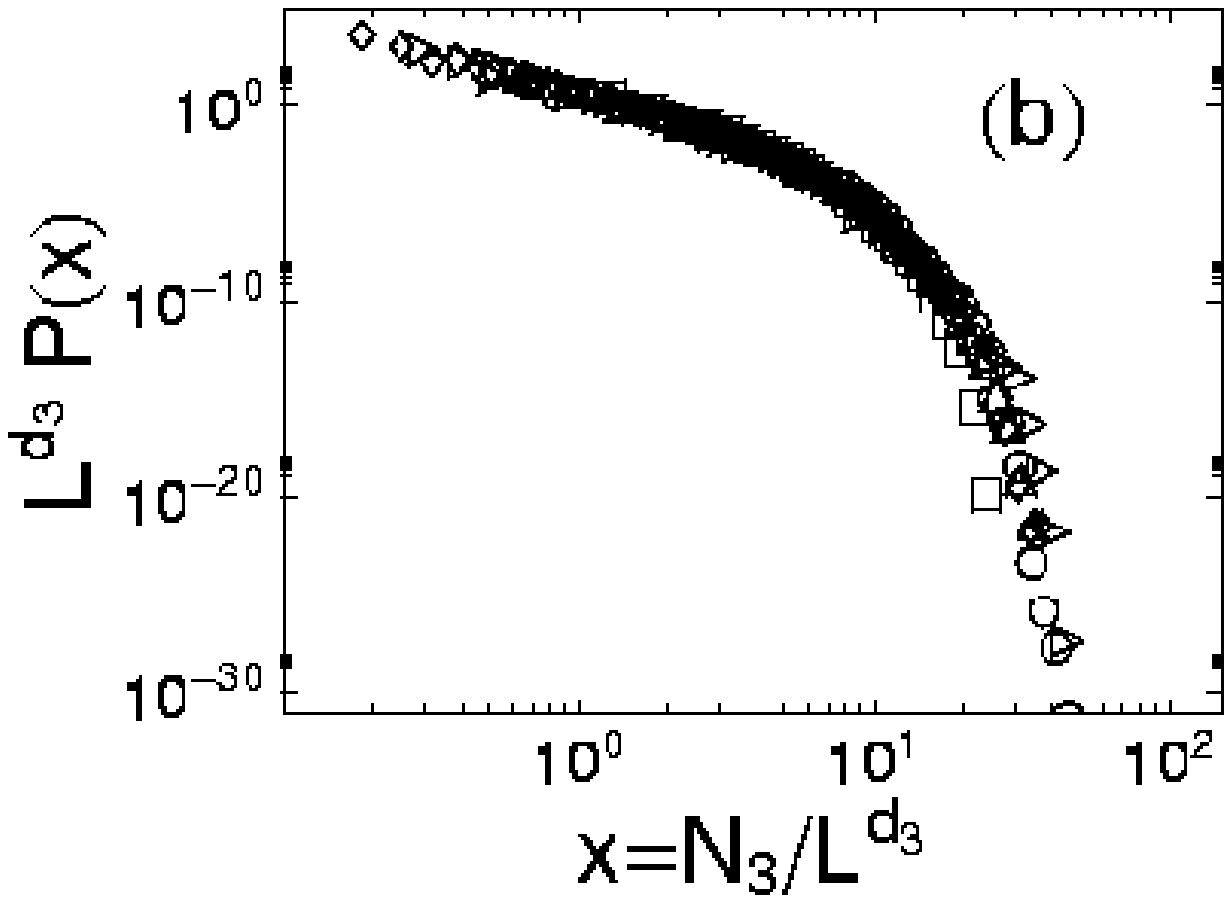}
}
\caption{$P(N_3|L)$, the distribution of 3-block mass $N_3$ for (from
left to right) $L=8$, 16, 32, 64, and 128 for the case of four
dimensions. (a) uncollapsed (b) collapsed using a value of $d_3=0.7$.}
\label{4d}
\end{figure}

\begin{figure}[tbh]
\centerline{
\xsize
\epsfclipon
\epsfbox{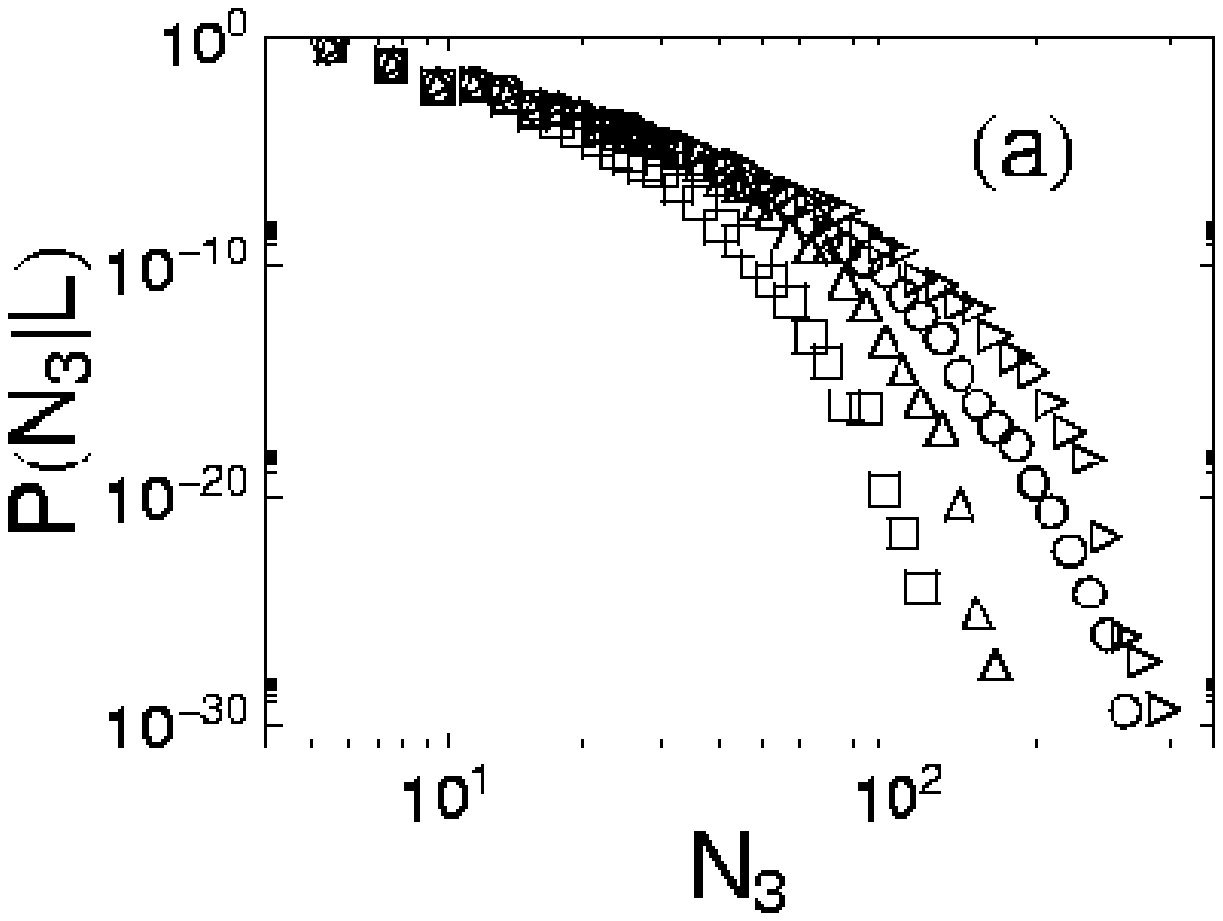}
}
\centerline{
\xsize
\epsfclipon
\epsfbox{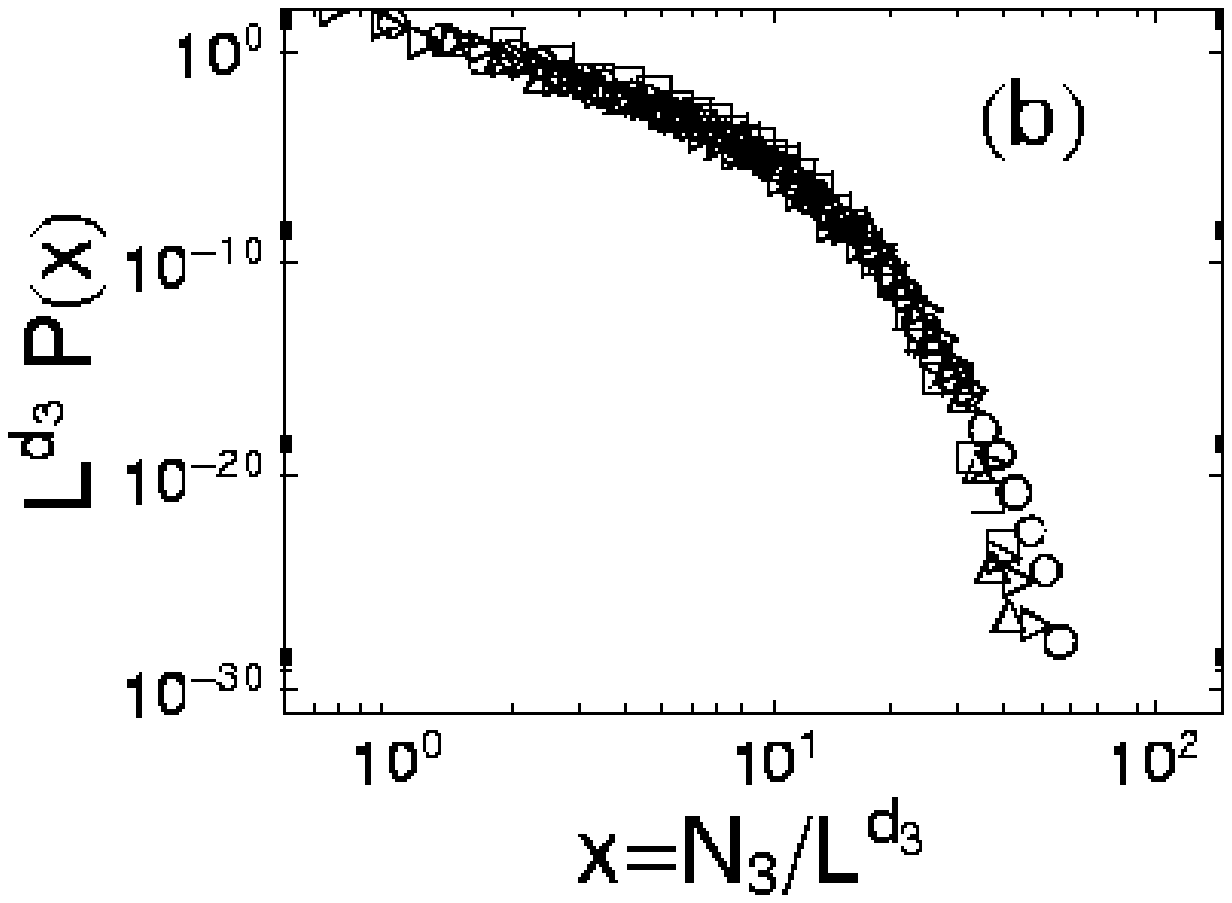}
}

\caption{$P(N_3|L)$, the distribution of 3-block mass $N_3$ for (from
left to right) $L=8$, 16, 32, and 64 for the case of five
dimensions. (a) uncollapsed (b) collapsed using a value of $d_3=0.5$.}
\label{5d}
\end{figure}

\begin{figure}[tbh]
\centerline{
\xsize
\epsfclipon
\epsfbox{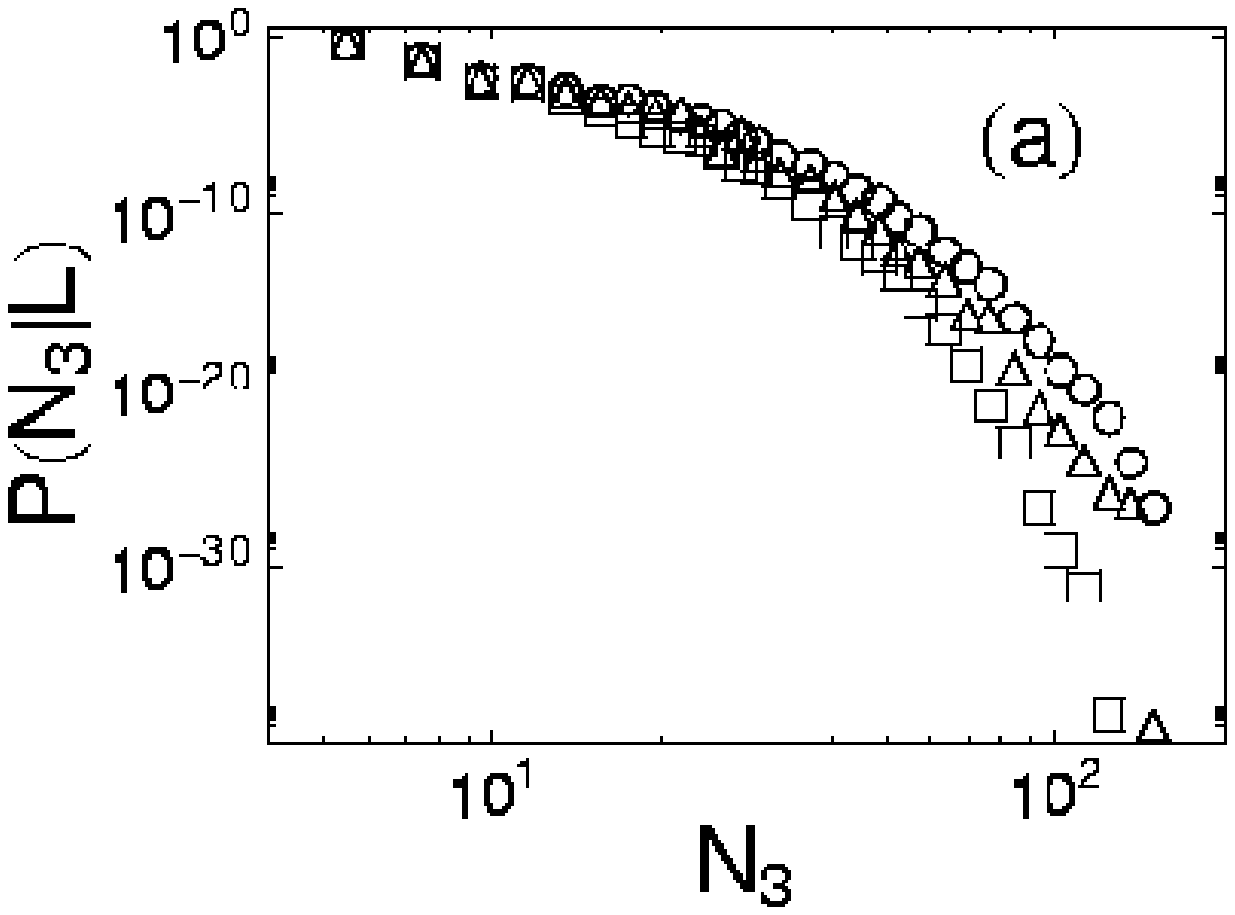}
}
\centerline{
\xsize
\epsfclipon
\epsfbox{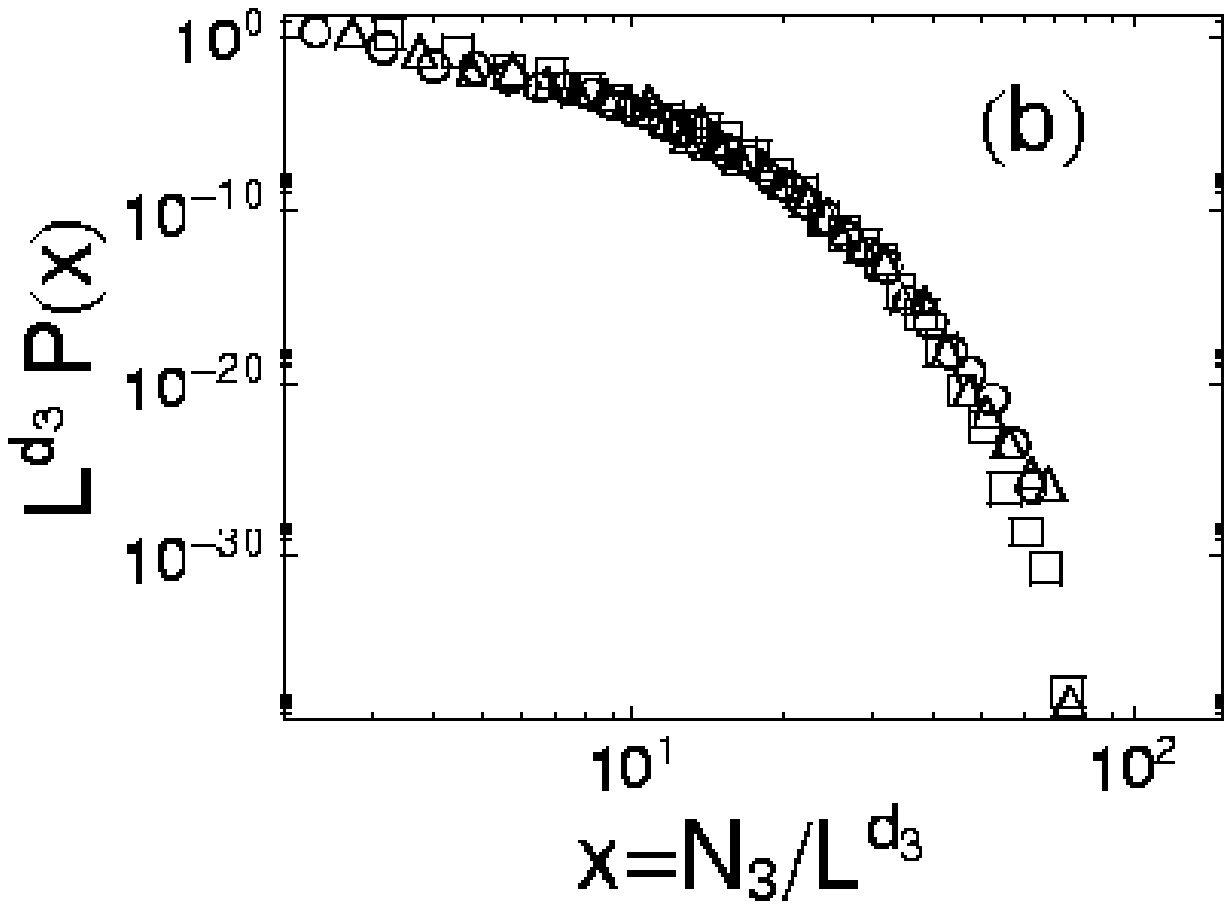}
}
\centerline{
\xsize
\epsfclipon
\epsfbox{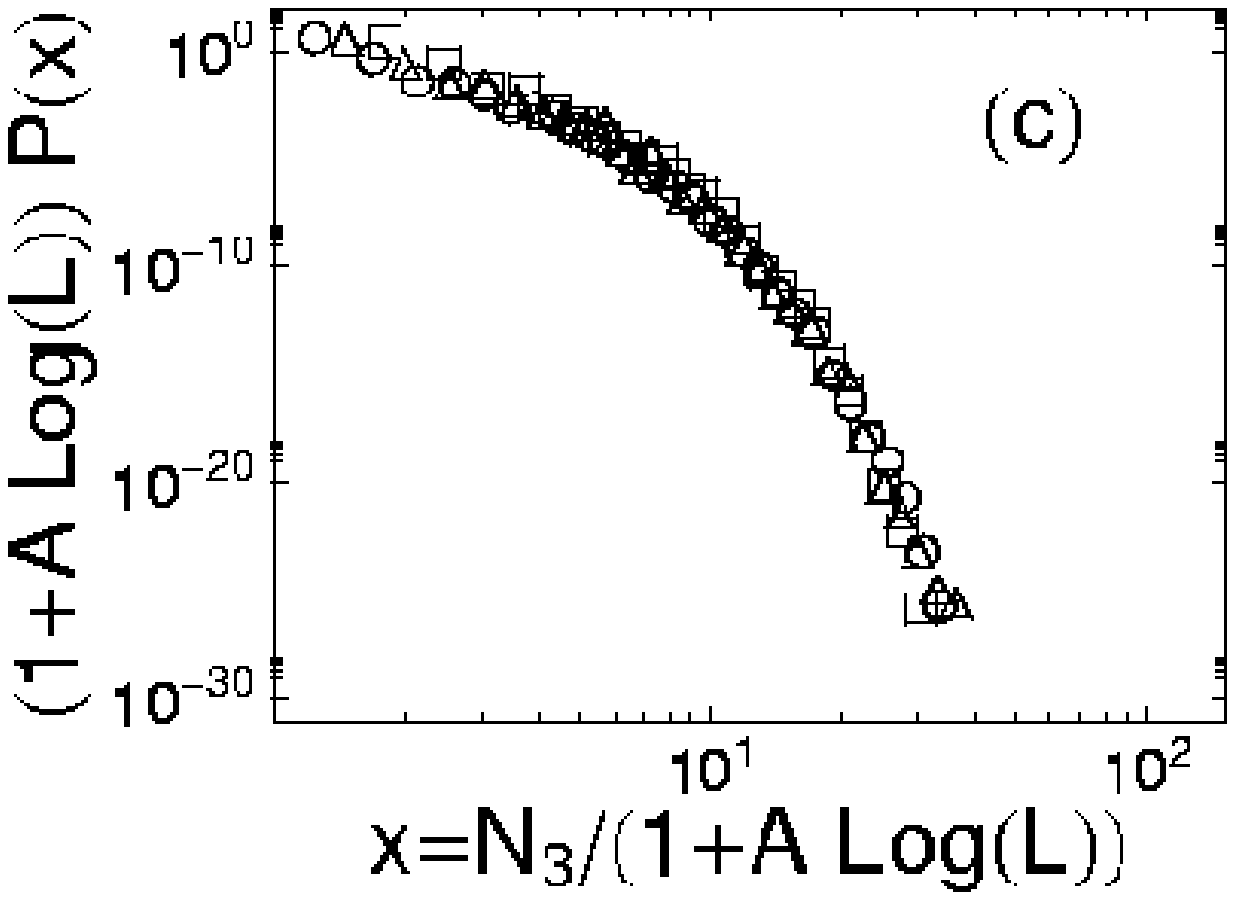}
}
\caption{$P(N_3|L)$, the distribution of 3-block mass $N_3$ for (from
left to right) $L=8$, 16, and 32 for the case of six dimensions. (a)
uncollapsed (b) collapsed using a value of $d_3=0.25$. (c) Collapsed
assuming $N_3\sim 1+A\log L$ with $A=1.0$.}
\label{6d}
\end{figure}

Cloning can take place multiple times during the growth of a cluster;
the result is a tree structure of realizations where the leaves of the
tree represent the completion of cluster growth.

Here $m$ and $n$ are parameters which can be tuned to achieve the
desired level of ``rareness'' which can be reached. If $n$ is large and
$m$ is small, there will be little cloning and we will generate clusters
with weights only moderately smaller than without enhancement. If $n$ is
small and $m$ is large, there will be much cloning and we will generate
clusters with weights very much smaller than without
enhancement. However, if $n$ is sufficiently small and/or $m$ is
sufficiently large, cluster growth will effectively never end naturally,
and we will not be able to extract useful information from the
simulation.

From an implementation standpoint, it is not necessary to actually
create copies of the state of the system in computer memory to create
the clones; as noted in Ref.~\cite{Grassberger} we can effectively walk
the clone tree in a ``depth-first'' manner, completely treating a given
clone before we begin treating the next clone. What is required is that
we save the state of the system before we begin growth based on a clone
so that we can return to that state when we begin growth on the next
clone. This saving of state is accomplished naturally with a
``last-in-first-out'' stack in which we maintain information about sites
in the cluster.

We first attempted to create realizations with large 3-blocks by
creating very dense clusters. We set as our criteria for cloning the
condition that the number of occupied bonds actually created during the
$n$ shell interval be larger than the number of times we determined
whether a bond should be occupied times the bond occupation
probability. While this algorithm is very effective in creating dense
clusters, it did not result in large 3-blocks within the clusters. We
were, however, successful in creating clusters with large 3-blocks by
using a criterion which results in the creation of large blobs: clone if
the most massive blob found in the cluster at the end of the interval
is more massive than the largest blob created before growth in the
interval is begun. That is, either an existing blob grows, one or more
blobs merge or a new blob is created which is more massive than any
existing previous to growth in the interval.

\subsection{Incremental Cluster Decomposition}

The decision whether to clone depends on a knowledge of the mass of the
largest blob in the cluster. It would be unacceptably inefficient to
decompose the entire cluster into blobs starting from scratch each time
we must make a cloning decision. Instead, we use a new algorithm for
cluster decomposition which allows us to incrementally decompose the
cluster into blobs. At the end of an interval of $n$ chemical shells of
growth, we need only consider the effect on the cluster decomposition of
the sites and bonds we have added to the cluster during the
interval. The algorithm, based on the algorithm of Ref.~\cite{Porto} for
determining the cluster backbone, works as follows:

\begin{itemize}

\item[{(i)}] During the growth of the cluster we identify ``loop
sites.'' Loop sites are sites which are reached from two or more
different growth sites simultaneously
\protect\cite{Porto,HermannandStanley}.

\item[{(ii)}] At the end of an interval of growth, we use the burning
algorithm \protect\cite{HermannandStanley,Porto} to walk back from each
loop site toward the origin of the cluster. When we reach a state during
the walk when only one site is burning, all sites traversed so far
compose a blob. If during the walk we hit an existing blob, that blob is
incorporated into the blob associated with the loop site from which
the walk started.

\item[{(iii)}] When we have exhausted all clones created at the end of
an interval, we must restore the system to its state at the beginning of
the interval. That is, we must: 
\begin{itemize}

\item[{(a)}] destroy all blobs created, 

\item[{(b)}] separate any blobs which were merged, and

\item[{(c)}] reduce any blobs which grew during the interval back to
their size at the beginning of the interval. 

\end{itemize}

This is all accomplished by carefully maintaining the appropriate state
information during the growth and cluster decomposition processes.

\end{itemize}

\section{Results and Discussion}
\label{Results}

Using the methods described in the previous section, we generate
percolation clusters on hypercubic lattices for 4d, 5d and 6d at their
respective percolation thresholds \cite{pzs,Grassbergerx}.

In Fig.~\ref{4d}a, we plot $P(N_3|L)$, the distribution of 3-block mass
$N_3$ in a system of size $L$ for various $L$ for $d=4$. In
Fig.~\ref{4d}b, we plot the same distributions collapsed using the
estimated value $d_3=0.7$ which, visually, yields the best collapse. We
show analogous plots for $d=5$ and $d=6$ Figs.~\ref{5d} and
\ref{6d}. Based on the value of $d_3$ which yields the best collapse, we
estimate
\begin{equation}
d_3 =
\begin{cases}
 0.7\pm 0.2 & \mbox{(4d)} \\
 0.5\pm 0.2 & \mbox{(5d)}.
\end{cases} \\
\end{equation} 
If we fit our results for 6d with a power law, then we find the best
collapse is obtained for $d_3=0.25\pm 0.2$. However it it difficult to
numerically distinguish between power-law scaling with a small exponent
and logarithmic scaling. Hence in Fig.~\ref{6d}c we also collapse the
distributions for 6d assuming $d_3=0$ with logarithmic corrections to
scaling
\begin{equation}
N_3\sim 1+A\log L \qquad \mbox{(6d)} 
\end{equation}
with $A=1.0$.  The quality of the collapses for power law scaling and
logarithmic scaling seem to be comparable; however, the facts that
$d_3=0$ for the Cayley tree and that logarithmic corrections to scaling
are common at the upper critical dimension favor the conclusion that
$d_3=0$ for $d=6$.

Finally, we make two observations:

(i) We note that the behavior of $d_3$ with dimension is qualitatively
the opposite of the behavior of $d_2$, the blob fractal dimension, in
the following sense: $d_2$ increases significantly between $d=2$ and
$d=3$ but increases very slowly between $d=3$ and $d=6$
\cite{Stauffer,Grassberger99,Moukarzel,Janssen}, while $d_3$ is slowly
decreasing between $d=2$ and $d=3$ but then decreases significantly
between $d=3$ and $d=6$.

(ii) Since $k=0$ corresponds to the entire system, which scales as
$L^d$, we note that for $k=$0, 1, 2, and 3, the fractal dimensions $d_k$
for 6d are 6, 4, 2, and 0 respectively, that is, a series of decreasing
consecutive even integers.

\begin{acknowledgments}

We thank Don Baker, Sergey Buldyrev, Shlomo Havlin, Greg Huber, and
Sameet Sreenivasan for helpful discussions and NSF, BP, and Intevep for
support.

\end{acknowledgments}

\end{document}